\documentclass[11pt,a4paper]{article}
\usepackage{jheppub_kim}
\usepackage{longtable}
\usepackage{epsfig}
\usepackage{amsmath}
\usepackage{graphicx}
\usepackage{hyperref}
\usepackage{epsfig,multicol,bbm}
\usepackage{lmodern}
\usepackage[T1]{fontenc}
 \usepackage{graphics}
 \usepackage[scriptsize,nooneline,hang]{caption}
\usepackage[hang,nooneline,scriptsize]{subfigure}
\usepackage{array}
\usepackage{makecell}
\usepackage{verbatim}

\usepackage[toc,page]{appendix}
\usepackage{graphicx}
\usepackage{colortbl}
\usepackage{ptext}
\usepackage{amsmath}
\usepackage{subfigure}
\usepackage{mathtools}
\usepackage[utf8]{inputenc}
\usepackage{float}
\usepackage[normalem]{ulem}
\usepackage{epstopdf}

\usepackage{amsfonts,amsmath,amssymb}
\usepackage{hyperref}

\usepackage{epsfig,multicol,bbm}
\usepackage{color}
\usepackage[dvipsnames]{xcolor}

\definecolor{darkblue}{rgb}{0.0, 0.0, 0.55}
\definecolor{grey}{rgb}{0.57, 0.64, 0.69}
\definecolor{lightbrown}{rgb}{0.71, 0.4, 0.11}
\newcommand{\tcb}{\textcolor{blue}}

\newcommand{\be}{\begin{equation}}
\newcommand{\ee}{\end{equation}}

\date{}

 \newcommand{\rbm}[1]{{\color{red}\bf [Robb: #1]}}

 \newcommand{\sns}[1]{{\color{blue}\bf [Naseh: #1]}}

\newcommand\fverb{\setbox\pippobox=\hbox\bgroup\verb}
\newcommand\fverbit{\egroup\item[\fbox{\unhbox\pippobox}]}

\newbox\pippobox

 \title{Black hole solutions to Einstein-Bel-Robinson gravity}

 \author[a]{S. N. Sajadi,}
  \author[b]{Robert B. Mann,}
  \author[c,d]{H. Sheikhahmadi,}
  \author[c]{M. Khademi}
\affiliation[a]{School of Physics, Institute for Research in Fundamental Sciences (IPM), \\P. O. Box 19395-5531, Tehran, Iran}
 \affiliation[b] {Department of Physics and Astronomy,
University of Waterloo, Waterloo, Ontario, N2L 3G1,Canada.}
 \affiliation[c]{School of Astronomy, Institute for Research in Fundamental Sciences (IPM),  \\P. O. Box 19395-5531, Tehran, Iran}
    \affiliation[d]{Center for Space Research, North-West University, Potchefstroom,  South Africa}
  
   \emailAdd{naseh.sajadi@gmail.com} \emailAdd{rbmann@uwaterloo.ca}
\emailAdd{h.sh.ahmadi@gmail.com;h.sheikhahmadi@ipm.ir}
    \emailAdd{maryam.khademi@ipm.ir}

\abstract{
\noindent In this paper, we study the physical properties of black holes in the framework of the recently proposed Einstien-Bel-Robinson gravity. We show that interestingly the theory propagates a transverse and massive graviton on a maximally symmetric background with positive energy. There is also a single ghost-free branch that returns to the Einstein case when $\beta\to 0$. We find new black hole solutions to the equations, both approximate and exact, the latter being a 
constant curvature black hole solution, and discuss inconsistencies with metrics that were previously claimed to be approximate solutions to the equations.
 We  obtain the conserved charges of the theory and briefly study the thermodynamics of the black hole solutions.}

\usepackage{mathrsfs}
\begin{document}
\maketitle
\flushbottom

\section{Introduction}\label{Intro}
\noindent Since more than a century ago, when modern gravity was introduced by Einstein, the nature of gravitation has   been   the focus of attention of physicists \cite{A.E.1915,A.E.1916}. And since the first days when Schwarzschild provided an exact solution for the  field equations of general relativity, the topic of black holes has moved in step with gravity \cite{Schw1916,Johannes1917}. 

The Einstein-Hilbert action, from which the field equations of general relativity follow, is  first order in the Ricci scalar. This formulation  has so far survived all empirical scrutiny to date. 
But topics such as the presence of quantum effects or higher dimensions have drawn attention to inclusion of higher orders of curvature in the action \cite{Smolic:2013gz} and have yielded interesting results in the context of  the renormalization of quantum gravity 
\cite{Stelle1977}. We generally expect such corrections to appear in the low energy limit of quantum gravity, with  Gauss-Bonnet corrections from string theory being a typical example \cite{Zwiebach1985,Cai2010}. Neither conformal fields   nor  holographic considerations have removed the motivation to consider such corrections, and indeed have   led to significant results in this field \cite{Myers-Robinson,Myers-etal,Myers-Sinha,Brigante-Robinson,Hennigar:2016gkm}, with quasi-topological gravity \cite{Myers-Robinson,Oliva2010} and generalized quasi-topological gravity being notable examples \cite{Hennigar:2016gkm,Bueno:2016xff}. 

 Despite the mathematical beauty and the computational complexity of  higher curvature gravity, in their general form they suffer from drawbacks \cite{Boulware1985}, particularly insofar as they contain ghosts in which the kinetic energy related to the emission of gravitons from a system becomes negative. This is an unacceptable phenomenon leading  to quantum unitarity breaking \cite{Hennigar:2016gkm}. Considerable efforts have been made to solve this ghost drawback. Lovelock gravity \cite{Lovelock} is the best known example, but it requires more than 4 spacetime dimensions.  Einsteinian cubic gravity is perhaps the simplest non-trivial generalization that is ghost-free on constant curvature backgrounds \cite{Hennigar:2016gkm,Bueno:2016xff}. 
 
 Modifications that are   quartic in the curvature \cite{quar1,quar2,Ahmed:2017jod,Ketov:2022lhx}  have also been of interest, primarily motivated by
ultraviolet renormalization of gravity  \cite{Smolic:2013gz}. Quadratic modifications were an early example along these lines \cite{Stelle1977,Stelle1978,Staro}.  Somewhat later string-theoretic considerations suggested that corrections that are at least quartic in curvature  are necessary \cite{Ketov2}.

In order to see the physical results of such quartic mathematical models, Einstein-Bel-Robinson (EBR) gravity \cite{Robinson} has recently been studied in a 4 dimensional context \cite{Ketov:2022lhx}.   A number of asymptotically flat black hole solutions to the field equations have been claimed \cite{CamposDelgado:2022sgc}, and these solutions have been used to study phenomenological corrections to general relativistic predictions.  These include light deflection and shadows \cite{Belhaj:2023dsn},  tidal corrections \cite{Arora:2023ijd}, slowly rotating black holes \cite{Davlataliev,Belhaj:2023GRG}, and quasinormal modes \cite{Bolokhov:2023dxq}, and shadows with nonzero cosmological constant \cite{Hamil:2023neq}.  Friedmann-Lemaitre-Robertson-Walker (FLRW) type solutions to the theory have also been considered  \cite{Ketov:2022zhp,Do:2023yvg}.
 
Here we point out that the  black hole solutions \cite{CamposDelgado:2022sgc} to EBR gravity do not solve its vacuum field equations.  Specifically we show the following.
\begin{itemize}
\item  Vacuum perturbative solutions in the EBR coupling $\beta$ require two metric functions under the assumption of spherical symmetry, {and both are required to obtain consistent solutions to the field equations. }
\item All existing static solutions in the literature assume identical metric functions and therefore do not satisfy the EBR field equations, even perturbatively.
 \item  Large-$r$ solutions asymptotic to the Schwarzschild-AdS solution also   exist, again requiring   two metric functions.
\end{itemize}
These results imply that the metrics originally considered in EBR gravity 
\cite{CamposDelgado:2022sgc,Belhaj:2023dsn,Arora:2023ijd,Davlataliev,Belhaj:2023GRG,Bolokhov:2023dxq,Hamil:2023neq,Mustafa:2024zsx}
must be accompanied by a non-zero stress energy that is engineered to ensure the field equations are satsified.

We also find black hole solutions to the field equations that are perturbative in the coupling $\beta$ -- these require two metric functions.  
Furthermore, we find that  a class of exact black hole solutions exist with negative constant curvature. 
These solutions are essentially the same as the well known solutions
in Einstein gravity  \cite{Aminneborg:1996iz,Mann:1996gj}, but with differing values of the effective cosmological constant.
 
Our main goal is to find   asymptotically AdS-black hole solutions to EBR gravity, which is a non-Lovelock, quartic-curvature theory in four dimensions. and to  briefly investigate their thermodynamic properties.  We begin by specifying the ghost-free branches of theory in Sec. \ref{sec2}.  
In Sec.~\ref{Sec3rad} we obtain solutions to the field equations at large-$r$, solutions perturbative in the coupling parameter  $\beta$,
 and exact black hole solutions of constant curvature.  In all cases but the last, two metric functions are required to consistently solve the equations.
 In Sec.~\ref{thermo} we consider the thermodynamic properties of these solutions. Finally Sec.~\ref{Conclusion} is devoted to concluding remarks and discussion. 


\section{4D Einstein-Bel-Robinson Gravity}\label{sec2}

In this section we will investigate the {basic features of EBR  theory including its basic field equations and the conditions required 
for freedom from ghosts.}

\subsection{Basic Action}\label{sub1}

In 4 dimensions, EBR gravity  is determined by the following action \cite{Ketov:2022lhx}
\begin{equation}\label{eq1}
\mathcal{S}=\dfrac{1}{16\pi G}\int d^{4}x\sqrt{-g}\left[R-2\Lambda -\beta \left(\mathcal{P}^2-\mathcal{G}^2\right)\right],
\end{equation}
where $R$ is the Ricci scalar, $\beta$ is the coupling constant of the theory, and $\Lambda=-3/\ell^2$, where $\ell$ refers to the curvature radius of the maximally symmetric
AdS solution of the field equations. The quantities $\mathcal{G}$ and $\mathcal{P}$ 
\begin{equation}
\mathcal{P}=\dfrac{1}{2}\sqrt{-g}\epsilon_{\mu \nu \rho \sigma}R^{\rho \sigma}{}_{\alpha \beta}R^{\mu \nu \alpha \beta}\;\;\;\;\;\;
\mathcal{G}=R^{2}-4R_{\mu\nu}R^{\mu\nu}+R_{\mu\nu\rho\sigma}R^{\mu\nu\rho\sigma}
\end{equation}
are the Euler and Pontryagin topological densities respectively, coming from the square of the Bel-Robinson tensor.
{Since all contributions from $\mathcal{P}$ vanish for spherical symmetry, we shall not consider it further.}

Varying the action \eqref{eq1} with respect to the metric yields  the following equations of motion
\begin{align}\label{fieldeq3}
\mathcal{E}_{a b}&\equiv R_{a b}-\dfrac{1}{2}g_{a b}R+\Lambda g_{a b} -\beta \mathcal{K}_{a b} = 0 \,,
\end{align}
where
\begin{equation}
	\begin{gathered}\label{Kab}
		\mathcal{K}_{a b}=\frac{1}{2} g_{ab} \mathcal{G}^2-2\Big[2 \mathcal{G} R R_{ab}-4 \mathcal{G} R_{a}^\rho R_{b \rho}
		+2 \mathcal{G} R_a^{\rho \sigma \lambda} R_{b \rho \sigma \lambda}+4 \mathcal{G} R^{\rho \sigma} R_{a \rho \sigma b}+2 g_{ab} R \square \mathcal{G}-\\
	2 R \nabla_a \nabla_b \mathcal{G}
		-4 R_{ab} \square \mathcal{G}+4\left(R_{a \rho} \nabla^\rho \nabla_b \mathcal{G}+R_{b \rho} \nabla^\rho \nabla_a \mathcal{G}\right) -4 g_{ab} R_{\rho \sigma} \nabla^\sigma \nabla^\rho \mathcal{G}+4 R_{a \rho b \sigma} \nabla^\sigma \nabla^\rho \mathcal{G}\Big]\,.
	\end{gathered}
\end{equation}

\subsection{Ghost Free Conditions}

We begin by assuming perturbative solutions exist about a constant curvature background. 
By decomposing the metric as
\begin{equation}
g_{ab}=\bar{g}_{ab}+h_{ab}\,,
\end{equation}
 one can  linearize the field equations \eqref{fieldeq3} around a locally AdS$_4$ vacuum to obtain
\begin{equation}\label{Backgg}
\bar{R}_{a b c d}=\dfrac{2}{3}\Lambda_{eff} \bar{g}_{a [c}\bar{g}_{b] d},\;\;\;\; \bar{R}_{a b}=\Lambda_{eff} \bar{g}_{a b},\;\;\;\; \bar{R}=4\Lambda_{eff}\,,
\end{equation}
where the background metric $\bar{g}_{ab}$ satisfies \eqref{Backgg}.
Regarding the perturbative behaviour, the linearized equation of motion can be expressed as
\begin{equation}\label{eqqlin}
\mathcal{E}^{(l)}_{a b}=R^{(l)}_{a b}-\dfrac{1}{2}\bar{g}_{a b}R^{(l)}-\beta \mathcal{K}_{a b}^{(l)}-\Lambda h_{ab}\,,
\end{equation}
where
\begin{align}
R^{(l)}_{a b}=&\dfrac{1}{2}\left(2h_{(a}{}^{c}{}_{;b);c}-h_{;b;a}-h_{a b}{}^{;c}{}_{;c}\right),\\
R^{(l)}=&h^{ac}{}_{;a;c}-h^{;a}{}_{;a}-\Lambda_{eff} h,\\
 \mathcal{K}_{a b}^{(l)}=&\dfrac{32\Lambda_{eff}^{2}}{9}\left[\bar{\nabla}_{a}\bar{\nabla}_{b}R^{(l)}-\bar{g}_{ab}\bar{\nabla}^{2}R^{(l)}-\Lambda_{eff} \bar{g}_{ab}R^{(l)}+{\dfrac{7}{9}\Lambda_{eff}^2 h_{ab}-\dfrac{4}{9}\bar{g}_{ab}\Lambda_{eff}^2 h}\right]\,,
\label{210}
\end{align}
where the superscript $^{(l)}$ is for the linearized (first order) perturbation and $h=\bar{g}^{ab}h_{a b}$ is the trace. Therefore the linearized field equation reads  
\begin{align}\label{eqqlin0}
{2\mathcal{E}^{(l)}_{a b}}&=2\Lambda h_{a b}-4\Lambda_{eff} h_{a b}+\Lambda_{eff} \bar{g}_{a b}h-h_{;a;b}+h_{b}{}^{c}{}_{;a;c}+h_{a}{}^{c}{}_{;b;c}-h_{ab}{}^{;c}{}_{;c}-\bar{g}_{ab}h^{cd}{}_{;c;d}+\bar{g}_{ab}h^{;d}{}_{;d}\nonumber\\
&-\dfrac{64}{81}\beta \Lambda_{eff}^{2}[19\Lambda_{eff}^{2}h_{ab}+2\Lambda_{eff}^{2}\bar{g}_{ab}h+21\Lambda_{eff} h_{;a;b}-21\Lambda_{eff} h_{b}{}^{c}{}_{;a;c}-27\Lambda_{eff} h_{a}{}^{c}{}_{;b;c}-\nonumber\\
&3\Lambda_{eff} \bar{g}_{ab}h^{cd}{}_{;c;d}+12\Lambda_{eff} \bar{g}_{ab}h^{;d}{}_{;d}+9h^{cd}{}_{;a;b;c;d}-9h_{;a;b;d}{}^{;d}-9\bar{g}_{ab}h^{cd}{}_{;c;d;e}{}^{;e}+9\bar{g}_{ab}h_{;d}{}^{;d;e}{}_{;e}]\,.
\end{align}
Now by imposing the transverse gauge, i.e., $\bar{\nabla}_{a}h^{ab}=\bar{\nabla}^{b}h$, 
and using 
\begin{align}
\bar{\nabla}_{c}\bar{\nabla}_{b}h^{c}{}_{a}&=\bar{\nabla}_{a}\bar{\nabla}_{b}h+\frac{4}{3}\Lambda_{eff}h_{ab}-\frac{1}{3}\Lambda_{eff}\bar{g}_{ab}h   \nonumber\\
\bar{\nabla}^{2}\bar{\nabla}_{a}\bar{\nabla}_{b}h&=\bar{\nabla}_{a}\bar{\nabla}_{b}
\bar{\nabla}^{2}h+\dfrac{8}{3}\Lambda_{eff}\bar{\nabla}_{b}\bar{\nabla}_{a}h-\dfrac{2}{3}\Lambda_{eff}g_{ab}\bar{\nabla}^{2}h\nonumber\\
\bar{\nabla}_{c}\bar{\nabla}_{d}\bar{\nabla}_{b}\bar{\nabla}_{a}h^{cd} &=
\bar{\nabla}_{b}\bar{\nabla}_{a}\bar{\nabla}^{2}h+4\Lambda_{eff}^{2}h_{ab}+\frac{14}{3}\Lambda_{eff}\bar{\nabla}_{a}\bar{\nabla}_{b}h-\Lambda_{eff}^{2}\bar{g}_{ab}h-\frac{2}{3}\Lambda_{eff}\bar{g}_{ab}\bar{\nabla}^{2}h \nonumber 
\end{align}
where the identity
$$
\bar{\nabla}_{c}\bar{\nabla}_{b}h_{a}^{c}-\bar{\nabla}_{b}\bar{\nabla}_{c}h_{a}^{c}=h_{a}^{c}\bar{R}_{bc}-h^{cd}\bar{R}_{acbd}
$$
was employed. In this gauge,  
the linearized field equations become 
\begin{align}\label{eqqlin}
&18\Lambda h_{a b}-12\Lambda_{eff}h_{ab}+3\Lambda_{eff}\bar{g}_{ab}h
+9\bar{\nabla}_{a}\bar{\nabla}_{b}h - 9\bar{\nabla}^{2}h_{ab} \nonumber\\
&-64\beta\Lambda_{eff}^2[-\Lambda_{eff}\bar{\nabla}_{a}\bar{\nabla}_{b}h +   \Lambda_{eff}\bar{g}_{ab}\bar{\nabla}^{2}h-
\Lambda_{eff}^{2}h_{ab}+\Lambda_{eff}^{2}\bar{g}_{ab}h]=0\
\end{align}
whose trace   yields
 \begin{equation}\label{hh-trace}
{\bar{\nabla}^{2}h-m_{s}^2h=0,\;\;\;\;\;m_{s}^2=-\Lambda_{eff}\left(1-\dfrac{3\Lambda}{32\beta \Lambda_{eff}^4}\right)}\,.
\end{equation}
For a ghost-free theory,  $m_{s}>0$, which   imposes  constraints on the parameters. 
Inserting the  constraint arising  from\eqref{hh-trace} into \eqref{eqqlin}, we find 
\begin{align}
{{\left(2\Lambda -\dfrac{4}{3}\Lambda_{eff}+\mathcal{A}\right)h_{ab}+\left(\dfrac{\Lambda_{eff}}{3}-\mathcal{B}\right)\bar{g}_{ab}h+
(1+\mathcal{C})\bar{\nabla}_{a}\bar{\nabla}_{b}h-\bar{\nabla}^{2}h_{ab}=0}}\,,
\end{align}
where
{
\begin{equation}
\mathcal{A}=\dfrac{64}{9}\beta\Lambda_{eff}^4,\;\;\;\mathcal{B}=\dfrac{64}{9}\beta\Lambda_{eff}^3m_{s}^2+\dfrac{64}{9}\beta\Lambda_{eff}^4,\;\;\;\mathcal{C}=\dfrac{64}{9}\beta\Lambda_{eff}^3\,.
\end{equation}
}
To extract the equation of motion for the traceless part 
of $h_{ab}$,  we write
\begin{equation}
h_{<ab>}\equiv h_{ab}-\dfrac{h}{4}\bar{g}_{ab}\,,
\end{equation}
and after some algebra obtain
\begin{align}\label{eqqtr1}
\left(2\Lambda-\dfrac{4}{3}\Lambda_{eff}+\mathcal{A}\right)h_{<ab>}+(1+\mathcal{C})\bar{\nabla}_{<a}\bar{\nabla}_{b>}h-
\bar{\nabla}^{2}h_{<ab>}=0\,.
\end{align}
Defining a new traceless tensor $\psi_{<ab>}$ via
\begin{equation}
h_{<ab>}=\psi_{<ab>}-\eta\bar{\nabla}_{<a}\bar{\nabla}_{b>}h
\end{equation}
we rewrite  \eqref{eqqtr1} and  choose $\eta$ such that the middle term in \eqref{eqqtr1} vanishes.  Using  
$$
[\bar{\nabla}^{2}\bar{\nabla}_{a}\bar{\nabla}_{b},\bar{\nabla}_{a}\bar{\nabla}_{b}\bar{\nabla}^{2}]h=\dfrac{8\Lambda_{eff}}{3}\bar{\nabla}_{a}\bar{\nabla}_{b}h-\dfrac{2\Lambda_{eff}}{3}\bar{g}_{ab}\bar{\nabla}^{2}h
$$
this gives 
\begin{equation}
\bar{\nabla}^{2}\psi_{<ab>}-\left({\dfrac{2}{3}\Lambda} +M^{2}\right)\psi_{<ab>}=0,\;\;\;\;\; \textrm{with}\qquad  M^2={\dfrac{4}{3}\Lambda} -\dfrac{4}{3}\Lambda_{eff}+\dfrac{64}{9}\beta\Lambda_{eff}^4\,,
\end{equation}
for the resulting equation of motion for the propagating massive mode $\psi_{<ab>}$. This implies that small excitations about a constant curvature background can be  associated with massive spin two particles (massive gravitons), where
\begin{equation}
{\eta=\dfrac{1+\mathcal{C}}{\mathcal{A}+2\Lambda -4\Lambda_{eff}-m_{s}^{2}}}\, .
\end{equation}
{The stability criterion in the AdS$_{4}$ background and freedom from ghosts respectively imply} 
 \begin{align}\label{LambdEff}
M^2>0\;\;\;\;\to\;\;\; &{\dfrac{64}{9}\beta \Lambda_{eff}^4-\dfrac{4}{3}\Lambda_{eff}+{\dfrac{4}{3}\Lambda} >0,}\\
m_{s}^2 {\geq 0} \;\;\;\;\;{\rightarrow}\;\;\;\; &{-\Lambda_{eff}\left(1-\dfrac{3\Lambda}{32\beta \Lambda_{eff}^4}\right)} {\geq 0}
\label{eqq2.22}
\end{align}
where $\Lambda_{eff}$ is chosen to satisfy  \eqref{LambdEff}. Requiring that the field equations admit constant curvature solutions yields
\begin{equation}
\dfrac{32}{9}\beta \Lambda_{eff}^4-\Lambda_{eff}+\Lambda =0 
\label{eqq2.23}
\end{equation}  
and insertion of this relation into \eqref{LambdEff} implies $\beta > 0$.
 
\begin{figure}[H]\hspace{0.4cm}
\centering
 \includegraphics[width=0.5\columnwidth]{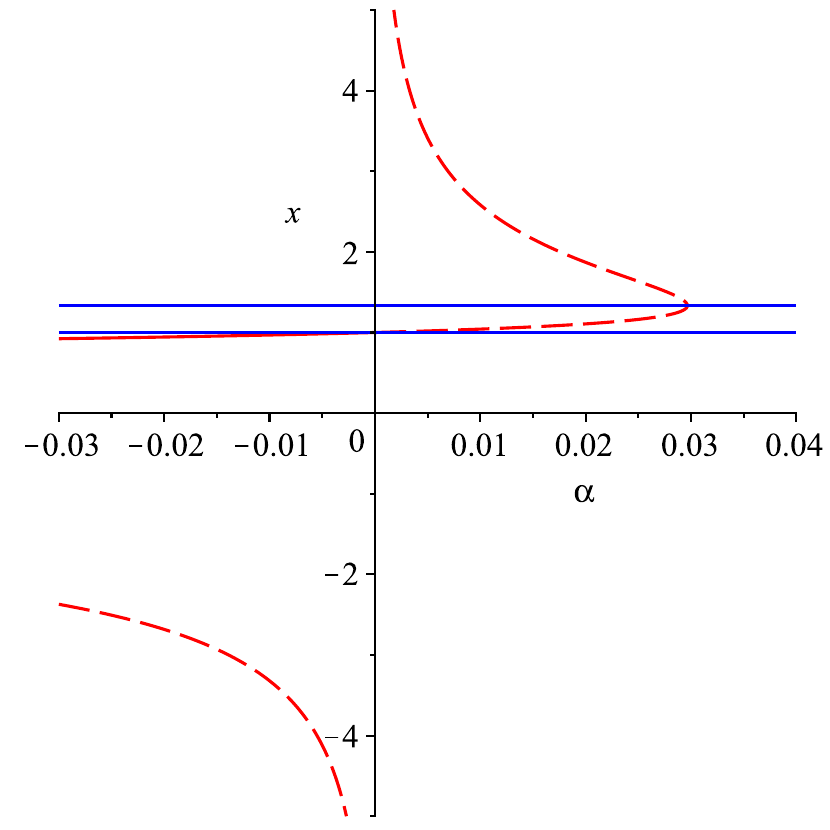}
\caption{ This plot illustrates the criteria  \eqref{eqq2.24} (red line) for constant curvature solutions to exist and the ghost-free constraint \eqref{eqq2.25}.
Solid lines are for $\Lambda <0$, in which case the regions outside the blue lines are permitted provided $x>0$; note that no solution for $x>1$ is allowed.
Dashed lines are for $\Lambda > 0$, {in which case the region ${x>0}$ is forbidden, except for in between the blue lines.}
 }
\label{TMSplote1}
\end{figure}

Rewriting these quantities in terms of $x=\Lambda_{eff}/\Lambda$ and $\alpha=\beta \Lambda^{3}$
we obtain
\begin{equation}
\dfrac{32}{9}\alpha x^{4}-x+1=0
\label{eqq2.24}
\end{equation}
from \eqref{eqq2.23} and
\begin{equation}
-\textrm{sgn}(\Lambda)x \left(1-\frac{1}{3(x-1)}\right) \geq 0
\label{eqq2.25}
\end{equation}
where the latter relation follows from inserting \eqref{eqq2.23} into \eqref{eqq2.22}. {Since $\beta > 0$, $\alpha$ and $\Lambda$ have the same sign.
The above relations, illustrated in figure~\ref{TMSplote1}, indicate that for $\Lambda<0$ ($\alpha<0$), $0<x<1$ is the allowed region for $x$, whereas
for $\Lambda>0$ ($\alpha>0$),  the allowed region is $-\infty<x<0$ and  $4/3<x<1$.}

\section{Radially Symmetric Metric  Solutions}\label{Sec3rad}

The most general radially symmetric metric can be written in the form
\be\label{sssmet}
ds^2 = -N(r) f(r) dt^2 + \frac{dr^2}{f(r)} + r^2 \left(d\theta^2 + \frac{\sin^2(\sqrt{k}\theta)}{k} d\phi^2\right)
\ee
where the metric functions $N(r)$ and $f(r)$ must be determined from the field equations  \eqref{fieldeq3}
and $k\in \lbrace 1,0,-1\rbrace$ respectively corresponding to spherical, planar, and hyperbolic transverse sections.  Note that in general $N$ is not constant, which is the generic situation for higher-curvature gravity theories.  The exception to this is the class of generalized quasitopological gravity theories (GQTGs) 
\cite{Hennigar:2016gkm,Bueno:2016xff,Bueno:2022res}, whose field equations ensure that $N$ is a constant that can be set to unity without loss of generality.  EBR theory is not in this class and
so $N$ will not be constant, as we shall see.  

 Obtaining exact solutions to these equations is in general prohibitively difficult.  We shall therefore consider various approaches toward finding approximate and perturbative solutions.  We then obtain an exact black hole solution of constant curvature.
 
  \subsection{Large-$r$ Solutions}
 
We begin by obtaining solutions that are asymptotic to constant curvature solutions.  To this end we write
 \be
 N(r) = 1 + \sum_{p=1} \frac{N_p}{r^p}  = 1 + \frac{N_1}{r} + \frac{N_2}{r^2} +\cdots \qquad  f(r) = -\frac{\Lambda_e}{3} r^2+k + \sum_{p=1} \frac{f_p}{r^p}  =  -\frac{\Lambda_e}{3} r^2+k+ \frac{f_1}{r} + \frac{f_2}{r^2} +\cdots 
 \ee
 where  if $\Lambda_e=0$ the solution is asymptotically flat. Inserting this into the $tt$ and $rr$ components of   \eqref{fieldeq3} yields
 \be\label{Nflgr}
  N(r) = 1 - \frac{252 ( \Lambda -\Lambda_e) m^2}{r^6 (14\Lambda - 15\Lambda_e) \Lambda_e^2} + \cdots
  \qquad 
  f(r) = -\frac{\Lambda_e}{3} r^2+k -\frac{2m}{r} {-}\frac{{84} ( \Lambda -\Lambda_e) m^2}{ r^4 (14\Lambda - 15\Lambda_e) {\Lambda_e}}   + \cdots
 \ee
 provided 
 \be \label{Keqn}
\frac{32}{9} \Lambda_e^4 \beta  -\Lambda_e + \Lambda = 0
 \ee
 which is equivalent to \eqref{eqq2.23} if $\Lambda_e = \Lambda_{eff}$.  It is possible to obtain exact solutions to \eqref{Keqn}, but they are not particularly enlightening.  Solving for $\Lambda_e$ in powers of $\beta$ yields
\be\label{eqq2.30}
\Lambda_e = \Lambda +  \frac{32 \Lambda^4}{9}\beta  + \frac{4096\Lambda^7}{81}\beta^2 + \cdots
\ee
which upon insertion into $f(r)$ in \eqref{Nflgr} is the same as the coefficient of the $r^2$ term in the perturbative solution \eqref{fbetsol} given below.

The leading terms   reproduce the (A)dS Schwarzschild solution. From these terms we see   that  the constant of integration $m$ corresponds to a mass parameter.  The remaining field equations are satisfied to order $1/r^4$.  Note that the quantity $N$ is not constant once the next order corrections to the 
(A)dS Schwarzschild solution are included.

\subsection{Perturbative Solutions in $\beta$}

Here we consider expansions of the form
\be\label{betapprox}
 N(r) = N_0 (r) + \beta N_1(r) + \beta^2 N_2(r) + \cdots \qquad f(r) = f_0 (r) + \beta f_1(r) + \beta^2 f_2(r) + \cdots
\ee
and attempt to solve \eqref{fieldeq3} perturbatively in $\beta$.  

Solving the $tt$ and $rr$ components of  \eqref{fieldeq3} yields
\begin{align}
N(r) &=  1 -\left(\frac{896 \Lambda m^{2}}{r^{6}}+\frac{3584 m^{3}}{r^{9}} \right) \beta \nonumber\\
&+\left(\frac{372736 \Lambda^{4} m^{2}}{9 r^{6}} - \frac{368640 k m^{2} \Lambda^{3}}{r^{8}} +\frac{10452992 m^{3} \Lambda^{3}}{9 r^{9}}
-\frac{17694720 k m^{3} \Lambda^{2}}{11 r^{11}} -\frac{8019968 m^{4} \Lambda^{2}}{12}
 \right. \nonumber\\
&\qquad  \left.  + \frac{50429952 k m^4 \Lambda}{r^{14}} -\frac{867926016  {m}^{5} \Lambda}{5 r^{15}} +\frac{4236115968  k m^{5}}{17 r^{17}}
-\frac{569393152 m^{6}}{r^{18}}
\right) \beta^2
\label{Nbetsol} \\
f(r) & =k-\frac{r^{2} \Lambda}{3}-\frac{2 m}{r} +\left( 
-\frac{32 r^{2} \Lambda^{4}}{27}-\frac{896 \Lambda^{2} m^{2}}{3 r^{4}}+\frac{768 \Lambda k m^{2}}{r^{6}}-\frac{3200 \Lambda m^{3}}{r^{7}}+\frac{4608 k m ^{3}}{r^{9}}-\frac{8576 {m}^{4}}{r^{10}}\right)\beta \nonumber\\
&\quad +\left(  -\frac{4096 r^{2} \Lambda^{7}}{243}+\frac{114688 \Lambda^{5} m^{2}}{9 r^{4}}-\frac{475136 \Lambda^{4} k^{2}}{3 r^{6}}+\frac{4280320 \Lambda^{4} m^{3}}{9 r^{7}}+\frac{327680 \Lambda^{3} k^{2} m^{2}}{r^{8}} \right. \nonumber\\
&\qquad  -\frac{2465792 \Lambda^{3} k m^{3}}{r^{9}}  -\frac{3162112 \Lambda^{3} m^{4}}{9 r^{10}} 
+\frac{1966080 \Lambda^{2} k^{2} m^{3}}{r^{11}}+\frac{219955200 \Lambda^{2} k^{4}}{11 r^{12}}
 \nonumber\\
& \qquad\quad 
-\frac{236548096 \Lambda^{2} m^{5}}{3 r^{13}} 
-\frac{47185920 \Lambda k^{2} m^{4}}{r^{14}} +\frac{3749331456 \Lambda k m^{5}}{r^{15}}-\frac{2936733696 \Lambda m^{6}}{5 r^{16}}  
 \nonumber\\
 &\left. 
\qquad\qquad -\frac{283115520 k^{2} m^{5}}{r^{17}}+\frac{20514373632 k m^{6}}{17 r^{18}}-\frac{1275707392 m^{7}}{r^{19}}
\right)\beta^2 
\label{fbetsol} 
\end{align}
where $m$ is a constant of integration.  It is straightforward to check that the remaining field equations are satisfied.  We see that 
$N$ is not constant even to linear order in $\beta$.
 The leading terms in \eqref{Nbetsol} and \eqref{fbetsol} reproduce the (A)dS Schwarzschild solution.  Again we see   from these leading order terms that $m$ corresponds to a mass parameter 

{Setting $\Lambda = 0$ (and $k=1$) yields perturbative asymptotically flat solutions. These solutions also have $N\neq 1$, even to linear order in $\beta$.
}

\subsection{Exact Black Hole Solution}\label{sec2p6}
 
 The field equations of the theory admit exact constant curvature solutions (pure AdS or pure dS).  Consequently we can obtain
 the solution 
 \begin{equation}\label{eqq3.1}
ds^{2}=-\left(\dfrac{\Lambda_e r^2}{3}-1\right)dt^{2}+\dfrac{dr^{2}}{\dfrac{\Lambda_e r^2}{3}-1}+r^2\left(d\theta^{2}+{\sinh^{2}(\theta)}d\phi^{2}\right)
\end{equation}
where 
\begin{equation}\label{new223}
\dfrac{32\beta  \Lambda_{e}^4}{9}+\Lambda_{e}+\Lambda=0
\end{equation}
which agrees with \eqref{eqq2.23} if $\Lambda_{e} = -\Lambda_{eff}$.  

The solution \eqref{eqq3.1} is a constant curvature black hole provided $\Lambda_{e} >0$. The rule of signs implies that for 
$\beta>0$ only $\Lambda <0$ yields a solution to \eqref{new223} that has $\Lambda_e > 0$, which in turn implies $\Lambda_{eff} < 0$. 
For $\beta<0$ there are at most two real positive solutions to \eqref{new223} if $\Lambda <0$, and one positive real solution if $\Lambda > 0$.

Writing  $\Lambda_e=3/L^2$, 
the horizon of this black hole is at $r_{+}= L = \sqrt{3/\Lambda_{e}} = \sqrt{-3/\Lambda_{eff}}$ and its mass  is zero.   Horizons of finite area can be obtained
by putting identifications in the transverse $(\theta,\phi)$ sections \cite{Aminneborg:1996iz,Mann:1996gj,Mann:1997iz}.

\section{Thermodynamics}
\label{thermo}

In this section we briefly examine the thermodynamic properties of the black hole solutions we have found.

To obtain the conserved charges we promote $\Lambda$ and $\beta$  to the scalar functions $\Lambda(x)$ and $\beta(x)$,
and couple these to field strengths $F_{\Lambda}(x)$ and $F_{\beta}(x)$ \cite{Hajian:2023bhq}.  This approach also allows us to
obtain the thermodynamic volume and to consider beta as a thermodynamic variable, along with their respective conjugates.
Details are given in appendix~\ref{appendixB}.

\subsection{Perturbative Solution}

For the perturbative solution \eqref{Nbetsol}, \eqref{fbetsol}
the temperature $T$ as a function of the horizon radius is 
\begin{align}
T=&\dfrac{1}{4\pi}\sqrt{\dfrac{1}{N}}[N(r)f(r)]^{\prime}\bigg|_{r_+} \nonumber\\
&=\dfrac{1}{4\pi r_{+}}-\dfrac{\Lambda r_{+}}{4\pi}+ 
\beta\left(-\dfrac{12\Lambda}{\pi r_{+}^{5}}+\dfrac{8\Lambda^2}{\pi r_{+}^{3}}-\dfrac{28\Lambda^3}{27\pi r_{+}}-\dfrac{2}{\pi r_{+}^{7}}-\dfrac{26\Lambda^4 r_{+}}{27\pi}\right)+\mathcal{O}(\beta^2).
\label{Tbetsol}
\end{align}
{From \eqref{Mcons},  the conserved mass $M=m$ to leading order in $\beta$.  Solving for $m$ by setting $f(r_+)=0$ and using the condition \eqref{eqq2.23}, 
we obtain}
\begin{equation}\label{eqM}
M= m+\mathcal{O}(\beta^2),
\end{equation}
where  
\begin{align}\label{eqm}
 {m} =\dfrac{r_{+}}{2}-\dfrac{\Lambda r_{+}^{3}}{6}+\beta\left(\dfrac{20}{r_{+}^{5}}-\dfrac{104\Lambda}{3r_{+}^{3}}+\dfrac{16\Lambda^2}{r_{+}}-\dfrac{56\Lambda^3 r_{+}}{27}-\dfrac{52\Lambda^{4}r_{+}^{3}}{81}\right)+\mathcal{O}(\beta^2).
\end{align}
 The exact expression for the entropy of the black hole is \cite{Wald1,Wald2}
\begin{equation}\label{eqqEnt}
S=-2\pi\int_{Horizon}d^{2}x\sqrt{\eta}\dfrac{\delta L}{\delta R_{a b c d}}\epsilon_{a b}\epsilon_{c d}=
\left(\dfrac{A}{4}\left[1-\dfrac{32\beta}{r^{4}}\left(2f^{\prime\prime}-f^{\prime 2}+\dfrac{3f^{\prime}N^{\prime}}{2N}\right)\right] \right)_{r_+}
\end{equation}
where 
\begin{align}\label{eqqdele}
\dfrac{\delta \mathcal{L}}{\delta R_{\mu \nu \rho \sigma}}=&
\dfrac{1}{16\pi G}\Big[\dfrac{1}{2}\left(g^{\mu \rho}g^{\nu \sigma}-g^{\mu \sigma}g^{\nu\rho}\right)+2\beta \mathcal{G}\Big[R(g^{\mu \rho}g^{\nu \sigma}-g^{\mu \sigma}g^{\nu\rho})\nonumber\\
&-4\left(g^{\mu \rho}R^{\nu \sigma}-R^{\mu \sigma}g^{\nu\rho}\right)+2R^{\mu\nu\rho\sigma}\Big]\Big]\,,
\end{align}
and
\begin{equation}
\eta= r_+^4\sin^{2}(\theta),\;\;\epsilon_{a b}=-2\sqrt{-\zeta}\delta^{t}_{[a}\delta^{r}_{b]},\;\zeta= -N(r_+)\,.
\end{equation}
To leading order in $\beta$
we obtain
\begin{equation}
S= \pi r_{+}^{2}+\dfrac{32\pi \beta}{r_{+}^{4}}\left(3-2\Lambda r_{+}^{2}+\Lambda^2 r_{+}^{4}\right)+\mathcal{O}(\beta^2) 
+ \cdots
\end{equation}

In the context of extended phase space thermodynamics, or what is more commonly known as black hole chemistry
\cite{Rob, Kubiznak:2012wp,Kubiznak:2016qmn} the cosmological constant can be interpreted as thermodynamic
pressure via
\be
P=-\frac{\Lambda}{8\pi}
\ee
with  conjugate volume 
\begin{equation}\label{pertvol}
V=\psi_{\Lambda}=\dfrac{4\pi r_{+}^{3}}{3}+\beta\left(\dfrac{448\pi}{3r_{+}^{3}}-\dfrac{704\pi\Lambda^2 r_{+}}{9}+\dfrac{1664\pi\Lambda^3 r_{+}^{3}}{81}\right)+\mathcal{O}(\beta^2)
\end{equation}
as shown in appendix \ref{appendixB}.

 We can also regard the coupling {$\frac{\beta}{16\pi} = B$ as} a thermodynamic parameter \cite{Gunasekaran:2012dq,Kastor:2010gq}, whose conjugate we
shall denote as $\psi_{\beta}$.  We find
\begin{eqnarray}
\psi_{\beta}&=&-\dfrac{64\pi}{r_{+}^{5}}-\dfrac{832\pi\Lambda^4 r_{+}^3}{81}+\dfrac{2560\pi\Lambda^3 r_{+}}{27}-\dfrac{128\pi\Lambda^2}{r_{+}}+\dfrac{256\pi\Lambda}{3r_{+}^{3}}  
\label{pertpsibet}\\
&+&\beta\left(\dfrac{143360\pi\Lambda^3}{9r_{+}^{5}}+\dfrac{3072\pi}{r_{+}^{11}} -\dfrac{23552\pi\Lambda^2}{r_{+}^{7}}+\dfrac{16384\pi\Lambda}{r_{+}^{9}}-\dfrac{99328\pi\Lambda^4}{27r_{+}^{3}}+\dfrac{13312\pi r_{+}\Lambda^6}{27}-\dfrac{4096\pi\Lambda^5}{9r_{+}}\right)+\mathcal{O}(\beta^2)
\nonumber
\end{eqnarray}
as also shown in appendix \ref{appendixB}.  

The first law of thermodynamics and Smarr formula are respectively expressed as follows 
\begin{equation}\label{1stlaw-Smarr1}
\delta M=T\delta S+\psi_{\Lambda}\delta P +\psi_{\beta}\delta B,\;\;\;\;\;\;M=2TS-2P\psi_{\Lambda}+6\psi_{\beta}B,
\end{equation}
and it is straightforward to show that both are satisfied to linear order in $\beta$.
 
The free energy is
\begin{align}
F= M - TS =\dfrac{r_{+}}{4}+\dfrac{\Lambda r_{+}^{3}}{12}+\beta\left(\dfrac{26\Lambda^{4}r_{+}^{3}}{81}+\dfrac{52\Lambda}{3r_{+}^{3}}-\dfrac{2}{r_{+}^{5}}+\dfrac{188\Lambda^{3}r_{+}}{27}-\dfrac{16\Lambda^2}{r_{+}}\right)+\mathcal{O}(\beta^2)
\end{align}
to leading order in $\beta$, and the  specific heat is 
\begin{align}
C=&T\left(\dfrac{\partial S}{\partial T}\right)_{P,B}=\dfrac{2\pi r_{+}^{2}(\Lambda r_{+}^{2}-1)}{1+\Lambda r_{+}^{2}}+\dfrac{32\pi \beta\left(-4\Lambda^{3}r_{+}^{6}+10\Lambda^2 r_{+}^{4}-12\Lambda r_{+}^{2}+9+\Lambda^{4} r_{+}^{8}\right)}{r_{+}^{4}(1+\Lambda r_{+}^{2})^{2}}+\mathcal{O}(\beta^2)\,.
\end{align}

\begin{figure}[H]\hspace{0.4cm}
\centering
 \includegraphics[width=0.75\columnwidth]{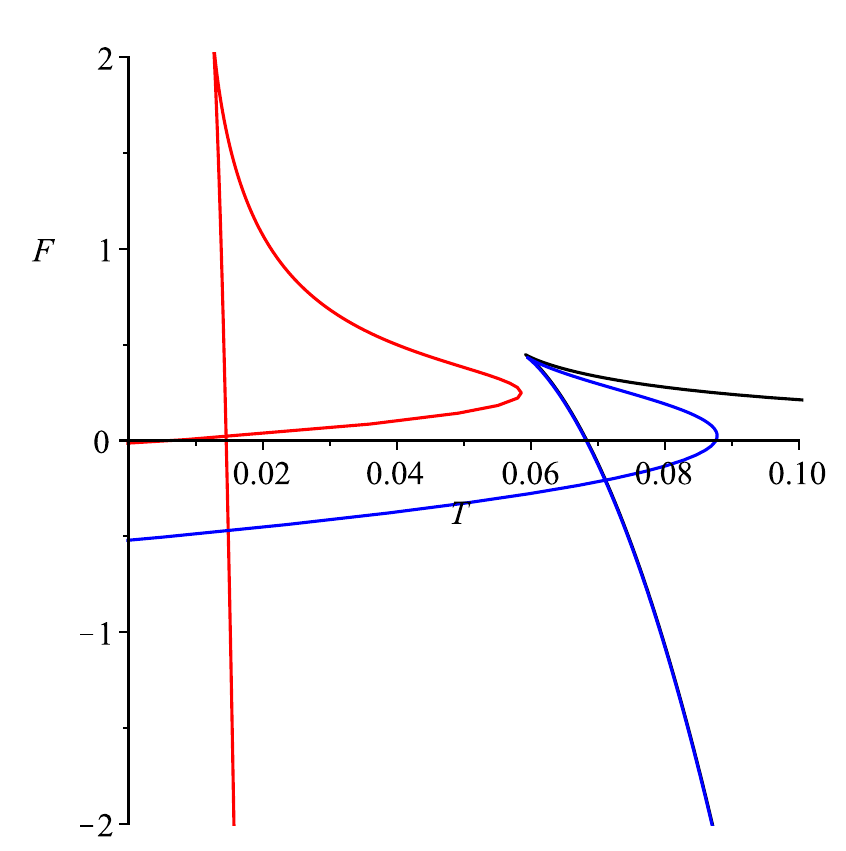}
\caption{ This figure shows the behaviour of free energy vs. temperature $T$ for $\beta=0.05,~P=\textcolor{red}{0.0025},\textcolor{blue}{0.0055}$;
The black curve depicts the free energy in Einstein gravity, with $\beta=0$ and $P={0.0055}$, which exhibits only a Hawking-Page transition.  
Nonzero values of $\beta$ admit first-order large-small black hole phase transitions (blue curve) as well as first-order transitions from a large 
black hole to radiation to a small black hole as the temperature decreases (red curve). All quantities are plotted in Planckian units.
}
 \label{FTplot}
\end{figure}

Without carrying out a full study of the the thermodynamics of EBR black holes, we plot in figure~\ref{FTplot} some of the possible behaviours that can emerge.
For $\beta=0$ we get the standard Hawking-Page transition from a large black hole to radiation as the temperature decreases, shown by the black curve.
Nonzero values of $\beta$ admit first-order large-small black hole phase transitions, with the signature swallowtail for the free energy, as shown in the blue curve
with $P=0.0025$; note that for these values of $(\beta,P)$ the free energy of the EBR large black hole is nearly indistinguishable from its counterpart in Einstein gravity.
However for smaller values of the pressure, more interesting phase behaviour can occur, as shown by the red curve with $P=0.0025$.  As temperature decreases, there is a
Hawking-Page transition from a large black hole to radiation; however a further decrease in temperature yields another first-order transition from radiation
to a small black hole, taking place at about $T=0.005$.  There will be a larger  value of the pressure (not shown in the figure) at which a large-small-radiation triple point will exist.  Such triple points have been seen before for exotic black holes in Lovelock gravity in higher dimensions \cite{Hull:2021bry}.  Here we see that such phenomena can occur in 4 spacetime dimensions.

A thorough study of EBR black hole thermodynamics remains an interesting subject for future investigation.

\subsection{Constant Curvature Solutions}

Writing $\Lambda_e = 3/L^2$ in \eqref{eqq3.1}, 
the temperature $T$ as a function of the horizon radius is 
\begin{equation}
T=\dfrac{f^{\prime}}{4\pi}\bigg|_{r_+} = \dfrac{r_{+}}{2\pi L^2}=\dfrac{1}{2\pi L}
\end{equation}

It is well-known from black hole chemistry that for A(dS) black holes one can take the pressure to be
\cite{Rob},\cite{Chernyavsky:2017xwm}
\begin{equation}
P=-\dfrac{\Lambda}{8\pi} = {\dfrac{1}{8\pi}\left(\dfrac{288 \beta}{L^8} + \dfrac{3}{L^2}\right)}
\end{equation}
from \eqref{new223}.  This  is  conjugate to the thermodynamic volume
{
\begin{equation}
V=\psi_{\Lambda}=
\omega \dfrac{L^9}{3(L^6+384\beta)} 
\end{equation}
where} $\omega = \int \sinh\theta d\theta d\phi$; this will be finite if the $(\theta,\phi)$ coordinates are appropriately identified \cite{Mann:1997iz}, in which case the horizon is compact and has a finite genus $g  \geq 2$.

 Using \eqref{eqqEnt} and \eqref{eqqdele}, the exact expression for the entropy of the black hole is \cite{Wald1,Wald2}
\begin{equation}
S= \omega\left[\dfrac{r_{+}^{2}}{4}+\dfrac{32\pi\beta}{L^{4}}\left(1-\dfrac{L^{2}}{r_{+}^{2}}\right)\right]_{r_{+}=L}=\dfrac{\omega\;L^2}{4}
\end{equation}
Following the approach in  appendix \ref{appendixB}, the thermodynamic potential corresponding to the coupling $B$ is
{
\begin{equation}
\psi_{\beta}=
-\dfrac{192\omega L}{L^6+384\beta}
\end{equation}
}
 
Finally we find that the first law of thermodynamics and Smarr formula  
\begin{equation}\label{1stlaw-Smarr}
0=T\delta S+ {\psi_{\Lambda}\delta P}+\dfrac{1}{16\pi}\psi_{\beta}\delta\beta,\;\;\;\;0=TS - {P\psi_{\Lambda}}+3B\psi_{\beta}\,.
\end{equation}
{are both satisfied exactly.}
 
The free energy of this solution is 
\begin{equation}
F = -TS =  -\dfrac{\omega\;L}{8\pi}
\end{equation}
in the fixed $(L,\beta)$ ensemble, where $L$ must satisfy \eqref{new223} upon setting $\Lambda_e=3/L^2$. For $\Lambda = -8\pi P$, this equation
has only one real solution for $L$ in terms of $P>0$, implying that the free energy $F$ is always a negative increasing function of the pressure $P$.

\section{Conclusion}\label{Conclusion}

Gravitational theories with curvature corrections of higher orders, in addition to their mathematical beauty and computational value, have found a significant place in high energy physics, particularly for  black holes.  We have studied the EBR model, which is a 4th order modification of  AdS Einstein-Hilbert gravity. 
By solving the  equations perturbatively about a constant-curvature background,  we have found a constraint that guarantees ghost-freedom for a range of values of the  parameter $\beta$. That such a  branch exists justifies the physicality of the model from the quantum point of view. 
 With the help of the Killing vector, we have also determined the conserved charges of the model, which play an important role in determining the energy of the system. 

 We showed that metrics that were previously claimed to be approximate solutions to the equations  are not approximate solutions.
By defining a general metric that has spherical symmetry, and is static, we obtained black hole solutions perturbative in $\beta$  to order $\beta^2$ and
an exact black hole solution of constant curvature.  All current static solutions in the literature 
\cite{CamposDelgado:2022sgc,Belhaj:2023dsn,Arora:2023ijd,Davlataliev,Belhaj:2023GRG,Bolokhov:2023dxq,Hamil:2023neq,Mustafa:2024zsx} do not satisfy the field equations, even perturbatively, since they assumed the metric function $N(r) = 1$ in the ansatz \eqref{sssmet}. Likewise,   large$-r$ solutions asymptotic to those in Einstein gravity (with or without cosmological constant) require two different metric functions.

We have also found an exact massless black hole solution of constant curvature, generalizing the topological black holes in Einstein-AdS gravity.  
Its free energy is a negative increasing function of the pressure in the fixed $(L,\beta)$ ensemble.

 For future work, in addition to the thermodynamics of EBR black holes, 
   the dynamical stability of the solutions merits study, for example via consideration of their  quasi-normal modes.  Slowly-rotating black hole solutions seem feasible to obtain and are likewise of interest, as would be a more thorough study of the thermodynamics of black holes in EBR gravity.
  
\section*{Acknowledgements}
This work was supported in part by the Natural Sciences and Engineering Research Council of Canada.  
  
\appendix

\section{Conserved Charges}
\label{conschg}

We next consider the conserved charges of the theory, aiming at studying the thermodynamics of the model properly. Any specific solution having a Killing vector $\bar{\xi}^{\mu}$ will have a corresponding conserved charge, obtained by contracting  this vector with both sides of \eqref{eqqlin}  \cite{Deser:2002jk}
 \begin{equation}
 Q^{a}(\bar{\xi})=Q^{a}_{E}(\bar{\xi})-\dfrac{32}{9}\beta\Lambda_{eff}^2 Q^{a}_{BR}(\bar{\xi})=\int d^{n-1}x\sqrt{-\bar{g}}\bar{\xi}_{b}T^{ab}=\int_{\Sigma}d\mathtt{S}_{i}F^{a i}\,,
 \end{equation}
 where the stress-energy tensor has been defined by setting ${\mathcal{E}^{(l)}_{a b}}=T_{ab}$, where  $F^{a i}$ is an anti-symmetric tensor that satisfies $T^{a b}\bar{\xi}_{b}=\bar{\nabla}_{b}F^{a b}$.
 The   terms $Q^{a}_{E}(\bar{\xi})$ and $ Q^{a}_{BR}(\bar{\xi})$ are respectively the $\beta=0$ term from Einstein gravity and
 the term from \eqref{eqqlin0} proportional to $\beta$. Explicitly we obtain \cite{Deser:2002jk}
 \begin{eqnarray}\label{QE}
 Q^{a}_{E}(\bar{\xi})&=&\int d\mathtt{S}_{i}\sqrt{-\bar{g}}\left[\bar{\xi}_{b}\bar{\nabla}^{a}h^{i b}-\bar{\xi}_{b}\bar{\nabla}^{i}h^{a b}+\bar{\xi}^{a}\bar{\nabla}^{i}h-\bar{\xi}^{i}\bar{\nabla}^{a}h+h^{a b}\bar{\nabla}^{i}\bar{\xi}_{b}-h^{i b}\bar{\nabla}^{a}\bar{\xi}_{b}
 + \right.  \nonumber\\
&&\left.
 \bar{\xi}^{i}\bar{\nabla}_{b}h^{a b}
 -\bar{\xi}^{a}\bar{\nabla}_{b}h^{i b}+h\bar{\nabla}^{a}\bar{\xi}^{i}\right]\,,
 \end{eqnarray}
and
\begin{eqnarray}\label{eqqcc}
 &&Q^{a}_{BR}(\bar{\xi})=\int d\mathtt{S}_{i}\sqrt{-\bar{g}}\left[R^{(l)}\bar{\nabla}^{i}\bar{\xi}^{a}+\bar{\xi}^{i}\bar{\nabla}^{a}R^{(l)}-\bar{\xi}^{a}\bar{\nabla}^{i}R^{(l)}\right]\,,
\end{eqnarray}
where this latter  contribution  comes from the tensor $\mathcal{K}_{ab}$ in \eqref{Kab}. This expression allows one to investigate  charges for solutions
that are deformations of a constant curvature background, a class of solutions that includes   static black holes.

{We now obtain the mass of the small $\beta$ black hole \eqref{Nbetsol} and \eqref{fbetsol} to leading order in $\beta$.}
We  assume the following expression 
\begin{equation}\label{BMetric}
ds^{2}_{b}=-\left(1-\dfrac{\Lambda_{eff}}{3}r^2\right)dt^2+\dfrac{dr^2}{1-\dfrac{\Lambda_{eff}}{3}r^2}+r^2d\theta^2+r^2\sin^{2}(\theta)d\phi^2\,,
\end{equation}
for  the background metric, about which we regard   \eqref{Nbetsol} and \eqref{fbetsol} as a perturbation. This gives
 \begin{align}
h_{tt}=&\dfrac{2m}{r}+\beta\left(\dfrac{640\Lambda m^{3}}{3r^{7}}+\dfrac{128\Lambda m^{2}}{r^{6}}+\dfrac{1408m^{4}}{r^{10}}-\dfrac{1024m^{3}}{r^{9}}\right)+\mathcal{O}(\beta^{2})\,,\\
h_{rr}=&\dfrac{486m}{\Lambda^{4}r^{9}}-\dfrac{108m^{2}}{\Lambda^{3}r^{8}}+\dfrac{18m}{\Lambda^{2}r^{5}}+\dfrac{108m}{\Lambda^{3}r^{7}}+\beta\left(-\dfrac{128m\Lambda}{r^{5}}-\dfrac{1152m}{r^{7}}-\dfrac{6912m}{r^{9}\Lambda}+\dfrac{3840m^2}{r^{8}}\right)+\mathcal{O}(\beta^{2})\,.
\end{align}
The energy associated with the future-directed timelike Killing vector $\xi^{\mu}= (-1, 0, 0 {,0})$, and using \eqref{eqq2.30} can be expressed as follows
\begin{equation}\label{Mcons}
{M =} E=Q^{(a)}(\bar{\xi})=m-\dfrac{1024\Lambda^{6}m}{27}\beta^2 +\mathcal{O}(\beta^3)
\end{equation}
{where we find that the $Q^{a}_{BR}(\bar{\xi})$ term makes no contribution in order $\beta$.}

{Now that we have calculated the conserved charges of the model, with the help of the above perturbative equations, we examine the thermodynamic properties of black hole solutions to the field equations.}

\section{Thermodynamic Potentials of Perturbative Solution}\label{appendixB}

To compute the relevant thermodynamic parameters we follow the approach in \cite{Hajian:2023bhq}, promoting
the couplings $\Lambda$ and $\beta$ are to scalars $\Lambda(x)$ and $\beta(x)$. We then pair these with 4-form field strengths 
$F_{\Lambda}(x) = dA_{\Lambda}$ and $F_{\beta}(x) = d A_{\beta}$.  The Lagrangian \eqref{eq1} then becomes
\begin{equation}
L=\dfrac{1}{16\pi}\left(R-2\Lambda(x)(1-F_{\Lambda}(x))+\beta(x)(\mathcal{G}^2-F_{\beta}(x))\right).
\end{equation}
Variation of the Lagrangian with respect to the new pairs of fields yields the  following on-shell relations:
\begin{equation}\label{Feqns}
F_{\Lambda}(x)=1,\;\;\;\;F_{\beta}(x)=\mathcal{G}^2 
\end{equation}
as well as
\begin{equation}
d{\Lambda}(x)=0 ,\;\;\;\; d{\beta}(x)= 0
\end{equation}
which ensures that $\Lambda(x)$ and $\beta(x)$ become coupling constants.

Equations \eqref{Feqns} have the solutions
 \begin{align}
\mathcal{F}_{\Lambda}(x)=&\sqrt{N(r)} r^2\sin(\theta) dt\wedge dr\wedge d\theta \wedge d\phi ,\\
\mathcal{F}_{\beta}(x)=&\dfrac{ \sin(\theta)}{\sqrt{N^7(r)} r^2}
\left[-4 N(r) f(r)(k-f(r))\left(\frac{\mathrm{d}^{2}}{\mathrm{~d} r^{2}} N(r)\right)-4 N(r)^{2}(k-f(r))\left(\frac{\mathrm{d}^{2}}{\mathrm{~d} r^{2}} f(r)\right)\right. 
\nonumber\\
&\qquad\qquad\qquad  +2 f(r)(k-f(r))\left(\frac{\mathrm{d}}{\mathrm{d} r} N(r)\right)^{2}-6\left(k-\frac{5 f(r)}{3}\right) N(r)\left(\frac{\mathrm{d}}{\mathrm{d} r} f(r)\right)\left(\frac{\mathrm{d}}{\mathrm{d} r} N(r)\right) \nonumber\\
& \left.  \qquad\qquad \qquad\qquad\qquad
+4 N(r)^{2}\left(\frac{\mathrm{d}}{\mathrm{d} r} f(r)\right)^{2}\right]^2 dt\wedge dr\wedge d\theta \wedge d\phi .
\end{align}
 The respective 3-form gauge potentials are {obtained by integrating over the radial variable}
\begin{eqnarray}
A_{\Lambda}(x)&=& -\left[\dfrac{r^3}{3}+\beta\left(\dfrac{224}{3r^{3}}-\dfrac{112\Lambda}{r}-\dfrac{112\Lambda^{3}r^{3}}{81}\right)+c_{1}\right]\sin(\theta)dt\wedge d\theta \wedge d\phi\\
A_{\beta}(x)
&=&-[\dfrac{144}{5r_{+}^{5}}-\dfrac{64\Lambda}{r_{+}^{3}}+\dfrac{160\Lambda^2}{r_{+}}-\dfrac{16r_{+}^{3}\Lambda^4}{3}+{64r_{+}\Lambda^3}+\beta(\dfrac{119808}{11r_{+}^{11}}+\dfrac{1288192\Lambda^4}{27r_{+}^{3}} \nonumber\\
&& -\dfrac{757504\Lambda^{5}}{27r_{+}}-\dfrac{132352\Lambda}{3r_{+}^{9}}-\dfrac{1067776\Lambda^3}{15 r_{+}^{5}}+\dfrac{509440\Lambda^2}{7r_{+}^{7}}-\dfrac{12032r_{+}^{3}\Lambda^7}{81}- \nonumber\\
&&\dfrac{20992r_{+}\Lambda^6}{27})+c_{2}] \sin(\theta)dt\wedge d\theta \wedge d\phi\;,
\end{eqnarray}
{to $1$-th order in $\beta$,} 
and $c_1$ and $c_2$ are arbitrary constants of integration. 
Evaluating these on the horizon, the corresponding thermodynamic conjugates are  
\begin{align}
\psi_{\Lambda}=\int \xi\;.\; A_{\Lambda}(x)
=&-\left[\dfrac{r_{+}^3}{3}+\beta\left(\dfrac{224}{3r_{+}^{3}}-\dfrac{112\Lambda}{r_{+}}-\dfrac{112\Lambda^{3}r_{+}^{3}}{81}\right)+c_{1}\right]\int \sin(\theta)d\theta \wedge d\phi\nonumber\\
=&-4\pi\left[\dfrac{r_{+}^3}{3}+\beta\left(\dfrac{224}{3r_{+}^{3}}-\dfrac{112\Lambda}{r_{+}}-\dfrac{112\Lambda^{3}r_{+}^{3}}{81}\right)+c_{1}\right],\label{eqqpsiL}
\end{align}
\begin{align}
\psi_{\beta}=&\int \xi\; .\; A_{\beta}(x) \nonumber\\
&=-4\pi\left[\dfrac{144}{5r_{+}^{5}}-\dfrac{64\Lambda}{r_{+}^{3}}+\dfrac{160\Lambda^2}{r_{+}}-\dfrac{16r_{+}^{3}\Lambda^4}{3}+{64r_{+}\Lambda^3}+\beta
\left(\dfrac{119808}{11r_{+}^{11}}+\dfrac{1288192\Lambda^4}{27r_{+}^{3}}\right.\right. \nonumber\\
& \left. \left. -\dfrac{757504\Lambda^{5}}{27r_{+}}-\dfrac{132352\Lambda}{3r_{+}^{9}}-\dfrac{1067776\Lambda^3}{15 r_{+}^{5}}+\dfrac{509440\Lambda^2}{7r_{+}^{7}}-\dfrac{12032r_{+}^{3}\Lambda^7}{81}- \dfrac{20992r_{+}\Lambda^6}{27}\right)+c_{2} \right],
\end{align}
 where we have used \eqref{eqm}.

Using the first law of thermodynamics, it is easy to fix the pure gauge part of the chemical potentials as follows:
First using \eqref{eqM} and \eqref{eqm} we have
\begin{align}
\delta_{r_{+}}M=&T\delta_{r_{+}}S.
\end{align}
It is easy to show that using $\delta_{r_{+}}\Lambda=0,\;\delta_{r_{+}}\beta=0$ and $\delta_{r_{+}}M=(\partial M/\partial r_{+})\delta r_{+}$ the above equality holds. Variation with respect to $\Lambda$ yields
\begin{align}
\delta_{\Lambda}M=&T\delta_{\Lambda}S+\dfrac{1}{8\pi}\psi_{\Lambda}\delta_{\Lambda}\Lambda+\dfrac{1}{16\pi}\psi_{\beta}\delta_{\Lambda}\beta,
\end{align}
since $\delta_{\Lambda}\beta=0,\;\delta_{\Lambda}\Lambda=1$ and using $\delta_{\Lambda}M=(\partial M/\partial\Lambda)\delta\Lambda$ then we have 
\begin{align}
&-\dfrac{r_{+}^{3}}{6}+\beta\left(-\dfrac{104}{3r_{+}^{3}}+\dfrac{32\Lambda}{r_{+}}-\dfrac{56r_{+}\Lambda^{2}}{9}-\dfrac{208\Lambda^{3}r_{+}^{3}}{81}\right)=\beta\left(-\dfrac{16}{r_{+}^{3}}+\dfrac{32\Lambda}{r_{+}}-16\Lambda^2 r_{+}\right)\nonumber\\
&-\dfrac{4\pi}{8\pi}\left[\dfrac{r_{+}^3}{3}+\beta\left(\dfrac{224}{3r_{+}^{3}}-\dfrac{112\Lambda}{r_{+}}-\dfrac{112\Lambda^{3}r_{+}^{3}}{81}\right)+c_{1}\right]
\end{align}
solving above for $c_{1}$ leads to 
\begin{equation}\label{eqqc1}
c_{1}=\beta\left(\dfrac{112}{3r_{+}^{3}}+\dfrac{176r_{+}\Lambda^{2}}{9}-\dfrac{112\Lambda}{r_{+}}-\dfrac{176\Lambda^3 r_{+}^{3}}{27}\right).
\end{equation}
and upon insertion of this  into \eqref{eqqpsiL}, we find
\begin{equation}
\psi_{\Lambda}=\dfrac{4\pi r_{+}^{3}}{3}+\beta\left(\dfrac{448\pi}{3r_{+}^{3}}-\dfrac{704\pi\Lambda^2 r_{+}}{9}+\dfrac{1664\pi\Lambda^3 r_{+}^{3}}{81}\right)+\mathcal{O}(\beta^2).
\end{equation}
Finally,  variation with respect to $\beta$ gives
\begin{align}
\delta_{\beta}M=&T\delta_{\beta}S+\dfrac{1}{8\pi}\psi_{\Lambda}\delta_{\beta}\Lambda+\dfrac{1}{16\pi}\psi_{\beta}\delta_{\beta}\beta,
\end{align}
using $\delta_{\beta}\Lambda=0$, $\delta_{\beta}\beta=1$ and $\delta_{\beta}M=(\partial M/\partial\beta)\delta\beta$ we have
\begin{eqnarray}
\psi_{\beta}&=&-\dfrac{64\pi}{r_{+}^{5}}-\dfrac{832\pi\Lambda^4 r_{+}^3}{81}+\dfrac{2560\pi\Lambda^3 r_{+}}{27}-\dfrac{128\pi\Lambda^2}{r_{+}}+\dfrac{256\pi\Lambda}{3r_{+}^{3}}  
+\beta\left(\dfrac{143360\pi\Lambda^3}{9r_{+}^{5}}+\dfrac{3072\pi}{r_{+}^{11}}\right. \nonumber\\
&&\left. -\dfrac{23552\pi\Lambda^2}{r_{+}^{7}}+\dfrac{16384\pi\Lambda}{r_{+}^{9}}-\dfrac{99328\pi\Lambda^4}{27r_{+}^{3}}+\dfrac{13312\pi r_{+}\Lambda^6}{27}-\dfrac{4096\pi\Lambda^5}{9r_{+}}\right)+\mathcal{O}(\beta^2)
\end{eqnarray}

\section{Thermodynamic Potentials of Exact Solution}
Equations \eqref{Feqns} for \eqref{eqq3.1} have the solutions
\begin{align}
\mathcal{F}_{\Lambda}(x)=& r^2\sinh(\theta) dt\wedge dr\wedge d\theta \wedge d\phi ,\\
\mathcal{F}_{\beta}(x)=&\dfrac{576r^{2}\sinh(\theta)}{L^{8}} dt\wedge dr\wedge d\theta \wedge d\phi .
\end{align}
The respective 3-form gauge potentials are {obtained by integrating over the radial variable}
\begin{eqnarray}
A_{\Lambda}(x)&=& -\left(\dfrac{L^3}{3}+c_{1}\right)\sinh(\theta)dt\wedge d\theta \wedge d\phi\\
A_{\beta}(x)
&=&-\left(\dfrac{192}{L^{5}}+c_{2}\right) \sinh(\theta)dt\wedge d\theta \wedge d\phi\;,
\end{eqnarray}
Evaluating these on the horizon, the corresponding thermodynamic conjugates are  
\begin{align}
\psi_{\Lambda}=&\int \xi\;.\; A_{\Lambda}(x)=-\left(\dfrac{L^3}{3}+c_{1}\right)\int \partial_{t}\;.\;dt\wedge d\theta \wedge d\phi \sinh(\theta)=-\left(\dfrac{L^3}{3}+c_{1}\right)\int \sinh(\theta)d\theta \wedge d\phi\nonumber\\
& =-\omega\left(\dfrac{L^3}{3}+c_{1}\right),\label{eqpsiL}\\
\psi_{\beta}=&\int \xi\;.\; A_{\beta}(x)=-\left(\dfrac{192}{L^{5}}+c_{2}\right)\int \partial_{t}\;.\;dt\wedge d\theta \wedge d\phi \sinh(\theta)=-\left(\dfrac{192}{L^{5}}+c_{2}\right)\int \sinh(\theta)d\theta \wedge d\phi\nonumber\\
& =-\omega\left(\dfrac{192}{L^{5}}+c_{2}\right)\label{eqpsiB}
\end{align}
Using the first law of thermodynamics \eqref{1stlaw-Smarr}, it is easy to fix the pure gauge part of the chemical
potentials as follows:
\begin{align}
T\delta_{L}S+\psi_{\Lambda}\delta_{L}P+\dfrac{1}{16\pi}\psi_{\beta}\delta_{L}\beta =0
\end{align}
since $\delta_{L}\beta=0$, then we have
\begin{align}
T\dfrac{\partial S}{\partial L}+\psi_{\Lambda}\dfrac{\partial P}{\partial L}=0\;\;\to\;\;\dfrac{\omega}{4\pi}-\dfrac{\omega}{8\pi}\left(\dfrac{L^3}{3}+c_{1}\right)\dfrac{\partial\Lambda}{\partial L}=0
\end{align}
{yielding 
\begin{equation}
c_{1}=-\dfrac{2 L^{3}(L^6+192\beta)}{3(L^6+384\beta)}
\end{equation}}
using \eqref{new223} with $\Lambda_3 = 3/L^2$. 
Inserting $c_{1}$ into \eqref{eqpsiL}, we obtain
\begin{equation}
\psi_{\Lambda}= \omega \dfrac{L^9}{3(L^6+384\beta)}
\end{equation}
Variation with respect to $\beta$ gives
\begin{align}
T\delta_{\beta}S+\psi_{\Lambda}\delta_{\beta}P+\dfrac{1}{16\pi}\psi_{\beta}\delta_{\beta}\beta =0
\end{align}
and since $\delta_{\beta}S=0$ and $\delta_{\beta}\beta=1$,   we have
\begin{equation}
\psi_{\Lambda}\dfrac{\partial P}{\partial\beta}+\dfrac{1}{16\pi}\psi_{\beta}=0\;\;\to\;\;
c_{2}=-\dfrac{73728\beta}{L^5(L^6+384\beta)}
\end{equation}
yielding
\begin{equation}
\psi_{\beta}=-\dfrac{192\omega L}{L^6+384\beta}
\end{equation}
for the thermodynamic conjugate to $B= 16\pi \beta$.

\end{document}